\documentclass[
reprint,
pre,
amsmath,amssymb,
aps,
floatfix,
longbibliography
]{revtex4-2}
\usepackage{mathtools}
\usepackage{extarrows}
\usepackage{graphicx}
\usepackage{dcolumn}
\usepackage[T1]{fontenc}

\usepackage{lineno}
\usepackage{multirow}
\usepackage{longtable}
\usepackage{setspace}
\usepackage{makecell}
\usepackage{setspace}
\usepackage{cancel}
\usepackage{enumitem} 
\usepackage{todonotes} %notes 
\usepackage{natbib}
\usepackage[normalem]{ulem}  
\usepackage{xcolor}  
\newcommand{\bb}{\mathbf}

\usepackage{booktabs}

\usepackage{xcolor}

\definecolor{myred}{RGB}{250,0,0}      % 暗红色
\definecolor{myblue}{RGB}{0,0,250}     % 深蓝色
\definecolor{myorange}{RGB}{250,100,0}  % 深橙色

\usepackage[final]{changes}

\newcommand{\add}[1]{#1}
\newcommand{\del}[1]{}

\newcommand{\dell}[1]{}

\newcommand{\delboth}[1]{}

\presetkeys{todonotes}{color=yellow!30,bordercolor=red,fancyline}{} 
\setcitestyle{numbers,square}
\usepackage[colorlinks,linkcolor=blue, urlcolor=blue,anchorcolor=blue, citecolor=blue]{hyperref}

\begin{document}
	%\linenumbers
	\preprint{}

\title{Space--time correlations of passive scalars in colored-noise flows}

\author{Long Wang}
\affiliation{State Key Laboratory of Nonlinear Mechanics, Institute of Mechanics,
Chinese Academy of Sciences, Beijing 100190, China}
\affiliation{School of Engineering Sciences, University of Chinese Academy of Sciences,
Beijing 100049, China}

\author{Guowei He}
\email{Contact author: hgw@lnm.imech.ac.cn}
\affiliation{State Key Laboratory of Nonlinear Mechanics, Institute of Mechanics,
Chinese Academy of Sciences, Beijing 100190, China}
\affiliation{School of Engineering Sciences, University of Chinese Academy of Sciences,
Beijing 100049, China}

\begin{abstract}
The space--time correlation of a passive scalar advected by a Gaussian colored-noise velocity with wavenumber-dependent correlation times and power-law spatial spectra is investigated in the present paper. Within the inertial-convective subrange, we derive an analytical solution for the space--time correlation. This solution validates the elliptic approximation (EA) model [He and Zhang, Phys. Rev. E 73, 055303(R) (2006)], demonstrating that the iso-correlation contours are self-similar in the co-moving space--time frame $(r-U\tau, V\tau)$, with a constant spatial-to-temporal intercept ratio of 1.55. Unlike the classic Kraichnan white-noise model, our formulation simultaneously recovers the Obukhov--Corrsin scaling for spatial correlations (when the velocity obeys Kolmogorov scaling) and reproduces the random-sweeping mechanism, yielding Gaussian (rather than exponential) temporal decorrelation of scalar Fourier modes. Our results clarify the underlying decorrelation mechanism of passive scalars: mean-flow advection and large-scale sweeping dominate temporal decorrelation, and small-scale distortion dominates spatial decorrelation. 

\end{abstract}          
	%\keywords{Suggested keywords}%Use showkeys class option if keyword
	%display desired
\maketitle
	%\tableofcontents
%%%%%%%%%%%%%%%%%%%%|Introduction Begin|%%%%%%%%%%%%%%%%%%%%
\section{Introduction}\label{sec:introduction}
Scalar mixing by turbulent flows, such as pollutant concentration or temperature fluctuations, is ubiquitous in natural and engineering systems, ranging from atmospheric dispersion and oceanic transport to industrial mixing processes~\cite{shraiman_scalar_Nature,falkovich2001particles,sreen_pnas_2019,warhaft_scalar_ARFM}. A scalar field is considered passive if its back-reaction on the flow is negligible. A central statistical quantity characterizing scalar transport is the space--time correlation, defined as the correlation of scalar fluctuations at two locations and two times, which characterizes the complex spatiotemporal structures of the scalar field \cite{he2017space,wallace2014space}. The temporal decorrelation of passive scalars is primarily governed by the random sweeping mechanism~\cite{kraichnan1964kolmogorov,chen1989sweeping}, whereby small-scale scalar structures are advected by large-scale energy-containing eddies, leading to a Gaussian decay of the temporal correlation of scalar modes. However, in the absence of an analytical derivation of this result from the fundamental scalar transport equation, a comprehensive understanding of space--time correlations remains elusive. In this paper, we will derive the scalar space--time correlation from the advection-diffusion equation with a colored-noise advecting velocity of finite correlation time. This analytical solution exhibits the Gaussian temporal decorrelation of scalar modes. Furthermore, we demonstrate that the scalar space--time correlation exhibits self-similarity in the inertial-convective subrange, consistent with the elliptic approximation (EA) model \cite{he2006elliptic,zhao2009space,he2010small}. The main contribution of the present work is to derive these results systematically from the scalar transport equation within a unified colored-noise framework.

Historically, closed-form analytical solutions for scalar correlations have stemmed from Kraichnan's white-noise model~\cite{kraichnan1968small} and random sweeping model~\cite{kraichnan1964kolmogorov,chen1989sweeping}. In the white-noise model, the velocity field is temporally white and spatially power-law correlated, leading to closed-form equations for multi-point scalar correlation functions. These equations show that zero modes determine anomalous scaling~\cite{kraichnan1994anomalous, Chertkov1995, falkovich2001particles}. The underlying dynamical mechanism for this intermittency can be traced to spontaneous stochasticity within the Lagrangian framework~\cite{drivas2017lagrangian, bernard1998slow, bandak2024spontaneous}. Remarkably, direct numerical simulations of the model faithfully reproduce the realistic ramp-cliff scalar structures observed in experiments~\cite{chen1998simulations, mestayer1976local, gibson1977structure, sreenivasan1979local, buaria2021small}. Furthermore, the Kraichnan white-noise model demonstrates that equal-time multiscaling does not necessarily imply dynamic multiscaling~\cite{Pandit_2005,ray2008universality}. Despite these considerable successes, the white-noise assumption leads to an exponential temporal decorrelation of scalar modes~\cite{Pagani2021, Gorbunova_2021, canet2022functional, PingFanYang2023}, which deviates from the Gaussian decorrelation observed in real turbulent flow~\cite{chen1989sweeping, sanada1992random, yeung2002random, Pullin2004modal}. In contrast, the random-sweeping model~\cite{kraichnan1964kolmogorov, chen1989sweeping} successfully captures the Gaussian decorrelation, yet relies on an ensemble of realizations in which the advecting velocity is held constant within each realization. These two models therefore represent two opposite extremes of vanishing and infinite velocity correlation times, respectively, whereas real turbulence lies between these two extreme cases~\cite{fannjiang2004, Boffetta2004, Sreenivasan2010}.

The finite correlation time of the velocity field has a profound impact on passive scalar transport. Boffetta \textit{et al.}~\cite{Boffetta2004} demonstrated that in finite-time correlated flows, the clustering of Lagrangian tracers remains qualitatively similar to that in the Kraichnan white-noise model but exhibits substantial quantitative deviations. Fannjiang~\cite{fannjiang2003,fannjiang2004} showed that for colored-noise flows with scale-dependent correlation times and power-law spatial spectra, the passive scalar converges to that of the Kraichnan model in the white-noise limit with higher spatial regularity. This regularity is related to Richardson's law of relative dispersion, whereby the relative diffusivity of tracer pairs scales as a $4/3$ power law in their separation. More importantly, particle-pair separation is no longer governed solely by local velocity increments but is significantly suppressed by the random sweeping of large-scale, energy-containing eddies~\cite{sokolov2000,chaves2003lagrangian,Thomson2005,eyink_PRE_2013}. Moreover, the zero-mode mechanism underlying anomalous scalar scaling has been established for exactly self-similar colored-noise velocity ensembles with a short but finite correlation time~\cite{Chertkov1996}. Although the advecting velocity in these studies is not taken directly from the Navier--Stokes equations, the colored-noise models nonetheless provide valuable physical insights into turbulent transport.

Theoretical attempt has been made to investigate the space--time correlations of passive scalars advected by velocity fields. Field-theoretic approaches, such as the functional renormalization group, predict a crossover from Gaussian to exponential behavior in the temporal correlation of scalar modes~\cite{canet2022functional}. However, direct numerical simulations only confirm the Gaussian decay~\cite{yeung2002random,Gorbunova_2021}. The EA model for the temperature field in turbulent Rayleigh--B\'enard convection~\cite{he2010small,ahlers2012logarithmic} implies that the space--time correlations are determined by the space correlations and two characteristic velocities: propagation and random-sweeping velocities. In spectral space, Wilczek and Narita~\cite{wilczek2012wave} developed a wavenumber-frequency model based on the random advection equation, which represents Doppler shifting and Doppler broadening. More recently, Wang and He~\cite{TK_wang2025} proposed the Taylor--Kraichnan (TK) model, in which the advecting velocity is the sum of a constant mean velocity, a random-sweeping velocity, and a white-noise velocity. The TK model admits an exact analytical solution for the space--time correlation and analytically confirms the EA model.

In this paper, we analytically derive the space–time correlations of passive scalars from a generalized Kraichnan model with colored-noise velocity. In this model, passive scalars are advected by synthetic velocity fields characterized by wavenumber-dependent temporal correlations and power-law spatial spectra~\cite{chaves2003lagrangian}. The results are used to elucidate the decorrelation process in scalar turbulence. The remainder of this paper is organized as follows. Section~\ref{sec:model} presents the model. This model is then used to derive the space--time correlations of passive scalars analytically in Section~\ref{sec:STC}. In Section~\ref{sec:EA}, the space--time correlations are specified in the inertial-convective subrange and used to illustrate the decorrelation process, consistent with the previous EA model. Section~\ref{sec:conclusion} is devoted to conclusions and discussion.

\section{Passive Scalar Model}\label{sec:model}
\subsection{Dynamics of scalar advection}
The dynamics of a passive scalar $\theta({\bf{x}},t)$ is described by the advection-diffusion equation, 
\begin{equation}\label{eq:dyn}
		\frac{{\partial \theta }}{{\partial t}} + (\bb{U}+{\bf{u}}) \cdot  \nabla  \theta  = \kappa { \nabla  ^2}\theta  + f,
\end{equation}
where the advecting velocity is composed of a mean flow $\bb{U}$ and a solenoidal fluctuating velocity $\bb{u}(\bb{x},t)$, $\kappa$ is the molecular diffusivity and $f$ an external injection source. The scalar field is sustained by a Gaussian, white-in-time source of zero mean, whose space--time correlation is prescribed as
\begin{equation}\label{eq:def-f}
\big\langle f(\bb{x},t) f(\bb{x}+{\bb r},t+\tau)\big\rangle
= \chi\left(\frac{r}{L}\right)\delta(\tau),
\end{equation}
where ${\bf r}$ is the spatial separation vector of magnitude $r$, and $\tau$ the time delay. $\chi$ sets the spatial amplitude of scalar injection at a large scale $L$. $\langle\cdot\rangle$ denotes an ensemble average, and $\delta(\tau)$ is the Dirac delta function~\cite{pope_2000}. In realistic turbulent flows, the advecting velocity field obeys the Navier--Stokes equations and manifests as colored noise with finite correlation time~\cite{Sreenivasan2010}. However, the Kraichnan model employs a synthetic velocity that is white in time, which facilitates analytical treatment but neglects temporal memory effects~\cite{shraiman_scalar_Nature,falkovich2001particles,Aiyer2017,Gorbunova_2021}.

For reference, the correlation functions, transport tensors, and scaling exponents used throughout this work are summarized in Table~\ref{tab:notation}.

\begin{table}[htbp]
\centering
\caption{Principal notation used in the present work.}
\label{tab:notation}
\begin{tabular}{ll}
\toprule
Symbol & Definition \\
\midrule
$D_{ij}(\mathbf r,\tau)$
& Fluctuating-velocity space--time correlation tensor \\

$d_{ij}(\mathbf r)$
& Equal-time velocity structure-function tensor \\

$S_{ij}(\mathbf r)$
& Two-particle relative-diffusivity tensor \\

$K_{ij}(\tau)$
& Single-particle diffusivity tensor \\

$C(\mathbf r,\tau)$
& Scalar space--time correlation \\

$C(\mathbf r,0)$
& Equal-time scalar correlation \\

$d_{\theta}(r,\tau)$
& Scalar space--time structure function \\

$\alpha$
& Velocity spatial-scaling exponent \\

$\beta$
& Velocity temporal-scaling exponent \\
\bottomrule
\end{tabular}
\end{table}

\subsection{Colored-noise velocity}

In the present work, we consider scalars passively advected by a colored-noise velocity field with power-law spatial spectra and wavenumber-dependent correlation times. The velocity field is Gaussian with a constant mean flow $\bb{U}=(U,0,0)$ in the streamwise direction, and the space--time correlation of the fluctuating velocity is~\cite{chaves2003lagrangian,Chatelain_2026}
\begin{equation}
\begin{aligned}\label{eq:def-model}
D_{ij}(\bb{r},\tau) &\equiv
\big\langle u_i(\mathbf{x},t)u_j(\mathbf{x+r},t+\tau)\big\rangle \\
&= D_2 \int \frac{{\rm e}^{-\tau D_3 k_L^{\beta}} P_{ij}(\mathbf{k}) {\rm e}^{{\rm i}\mathbf{k}\cdot\mathbf{r}}}
{k_L^{3+\alpha}}\frac{d\mathbf{k}}{(2\pi)^3},
\end{aligned}
\end{equation}
where \( P_{ij}(\mathbf{k}) = \delta_{ij} - \frac{k_i k_j}{k^2} \) is the transverse projector, and \( k_L = \sqrt{k^2 + L^{-2}} \) introduces an infrared cutoff at the large scale $L$. The exponent \( 0 < \alpha < 2 \) is the spatial H\"older (roughness) exponent of the velocity field, while the exponent $\beta > 0 $ governs the temporal decorrelation. The constants \( D_2 \) and \( D_3 \) are dimensional coefficients with units of \(length^{2-\alpha}/ time^2\) and \({ length}^{\beta}/{time}\), respectively. The equal-time spatial correlation is the same as that of the classical Kraichnan model, i.e., in the limit of $L \rightarrow \infty$,
\begin{align}
D_{ij}(\bb{r},0) = D_0^{(\alpha)}\delta_{ij}-d_{ij}^{(\alpha)}(\bb{r}),
\end{align}
\begin{align}\label{eq:structure}
d_{ij}^{(\alpha)}(\bb{r})= {D_1^{(\alpha)}}\left( (2+\alpha)\delta_{ij}-\alpha\frac{r_i r_j}{r^2}\right)r^{\alpha},
\end{align}
where $d_{ij}^{(\alpha)}(r)$ is the spatial structure function. $D_0^{(\alpha)}$ and $D_1^{(\alpha)}$ are constants corresponding to the scaling exponent $\alpha$~\cite{eyink2000self},
\begin{align}\label{eq:D_0}
D_0^{(\alpha)}&=\frac{2}{3}D_2\frac{\Gamma \left(\frac{\alpha}{2}\right)\pi^{d/2}}{\Gamma\left(\frac{\alpha+d}{2}\right)}L^{\alpha}, \\
D_1^{(\alpha)}&=\frac{1}{3+\alpha} D_2\frac{\Gamma(1-\frac{\alpha}{2})}{2^{\alpha} \cdot \alpha \cdot \Gamma\left(\frac{3+\alpha}{2}\right)}.
\end{align}
Note that $D_0^{(\alpha)}=D_{11}(\bb{0},0)=V^2$, where $V$ is the root-mean-square (r.m.s.) velocity component along any coordinate axis. 

The exponential temporal correlation in Eq.~\eqref{eq:def-model} can be replaced by any integrable function $\rho(D_3 k_L^\beta \tau)$ that decays to zero as $\tau \to \infty$~\cite{fannjiang2003}. Since the exponential form agrees well with the Lagrangian correlation measurements~\cite{Sato_Yamamoto_1987,Yeung_Pope_1989}, we use it here for simplicity. The characteristic correlation time scales as $\tau_c(r) \propto D_3^{-1}r^{\beta} $. Consequently, this finite-time-correlated model reduces to the Kraichnan model in the limit of $D_3 \rightarrow \infty$ with $D_2/D_3$ held fixed and to the frozen flow limit as $D_3 \rightarrow 0$~\cite{chaves2003lagrangian}.

The present paper is confined to the inertial-convective subrange. Within this regime, the velocity correlation \eqref{eq:def-model} is assumed to follow Kolmogorov scaling with $\alpha = \beta = 2/3$.

\section{Space--Time Correlations of Passive Scalars}\label{sec:STC}
\subsection{Space--time correlation}
The space--time correlation of a passive scalar $\theta$ is defined as
\begin{equation}\label{eq:def_STC}
C({\bf r},\tau) = \big\langle \theta({\bf x},t)\, \theta({\bf x}+{\bf r},t+\tau) \big\rangle,
\end{equation}
where $\bb{r}=(r,0,0)$ is the streamwise spatial separation and $\tau$ the temporal separation. Starting from the advection--diffusion equation and applying the
Furutsu--Novikov--Donsker (FND) theorem~\cite{frisch1995turbulence}
for the Gaussian velocity field, we perform a short-memory expansion of the recent-response contribution
over the interval $[t,t+\tau]$, valid for $\tau\ll t_r$, where $t_r$
is the turnover time at scale $r$. Retaining the zeroth-order
recent response yields the following approximate closed space--time correlation equation
(see Appendix~\ref{app:STC} for details):
\begin{equation}\label{eq:STC-pde}
\frac{\partial C({\bf r},\tau)}{\partial \tau}
=
-U_i\frac{\partial C({\bf r},\tau)}{\partial r_i}
+\frac{\partial}{\partial r_i}
\left[
K_{ij}(\tau)
\frac{\partial C({\bf r},\tau)}{\partial r_j}
\right],
\end{equation}
where the effective one-particle diffusivity tensor is approximated by
\begin{equation}
K_{ij}(\tau)
\simeq
\kappa\delta_{ij}+K_{ij}^{(0)}(\tau),
\end{equation}
and the zeroth-order turbulent contribution is
\begin{equation}
K_{ij}^{(0)}(\tau)
=
\int_0^\tau \mathrm{d}\sigma\,
D_{ij}(\mathbf 0,\sigma).
\end{equation}

For isotropic velocity fluctuations, Eq.~\eqref{eq:STC-pde} reduces to
\begin{equation}\label{eq:STC_radial_eq}
\frac{\partial C(r,\tau)}{\partial \tau} 
= -U\dfrac{\partial C(r,\tau)}{\partial r_1}+\frac{1}{r^2} \dfrac{\partial}{\partial r} 
  \left( r^2 K_{\parallel}(\tau)
  \dfrac{\partial C(r,\tau)}{\partial r} \right),
\end{equation}
where the effective longitudinal diffusivity is
\begin{equation}\label{eq:K_ll}
K_{\parallel}(\tau)
\simeq
\begin{cases} 
D_0^{(\alpha)}\tau+\kappa=V^2\tau+\kappa,
& \tau \ll \tau_c(L) , \\[6pt]
D_3^{-1}D_0^{(\zeta)}+\kappa,
& \tau \gg \tau_c(L) ,
\end{cases}
\end{equation}
with  
$\zeta=\alpha+\beta$, and the two asymptotic forms follow from the velocity covariance
model in Eq.~\eqref{eq:def-model}.
We consider a high-P\'eclet-number flow with a large integral scale $L$, yielding a broad inertial range in which molecular diffusion is negligible compared with turbulent transport. For time lags within the inertial range, \(\tau \ll \tau_c(L)\sim D_3^{-1}L^\beta\), so that the longitudinal diffusivity becomes
\begin{equation}\label{eq:Kll_approx}
K_{\parallel}(\tau) \simeq D_0^{(\alpha)} \tau = V^2 \tau.
\end{equation}

By solving Eqs.~\eqref{eq:STC_radial_eq} and \eqref{eq:Kll_approx}, we find that the space--time correlation is a convolution of the equal-time spatial correlation and the Gaussian displacement distribution:
\begin{equation}\label{eq:R=R_sP}
C(\mathbf{r},\tau)=\int \mathrm{d}\boldsymbol{\sigma}\,C(\mathbf{r}-\boldsymbol{\sigma},0)\,\frac{1}{(2\pi V^2\tau^2)^{3/2}}\exp\!\left[-\frac{\|\boldsymbol{\sigma}-\mathbf{U}\tau\|^2}{2V^2\tau^2}\right].
\end{equation}
This result is consistent with the Corrsin--Kovasznay conjecture~\cite{Corrsin1951,kovasznay1953turbulence} for space--time correlations. In spectral space, this Gaussian convolution yields the time correlation of the scalar modes, 
\begin{equation}\label{eq:E(k,tau)}
\begin{aligned}
E(\mathbf{k},\tau)&\equiv\langle \hat{\theta}(\mathbf{k},t) \hat{\theta}(-\mathbf{k},t+\tau)\rangle \\&= E(\mathbf{k})\exp \left(-\frac{1}{2}k^2V^2\tau^2-\mathrm{i}\bb{k}\cdot\bb{U}\tau\right),
\end{aligned}
\end{equation}
where $k=\|\bb{k}\|$. The real part of the exponent represents the Gaussian decay due to random sweeping, and the imaginary part represents the linear phase shift induced by the constant mean velocity. 

In the absence of a mean flow, Eq.~\eqref{eq:E(k,tau)} reduces to Gaussian temporal decorrelation, in agreement with direct numerical simulations of passive scalar transport in homogeneous isotropic turbulence~\cite{Gorbunova_2021}. For a nonzero mean velocity, it recovers the functional form of the Wilczek--Narita model~\cite{wilczek2012wave}, originally proposed for velocity fields and later validated in channel-flow turbulence~\cite{wilczek2015spatio}. Taking the Fourier transform of Eq.~\eqref{eq:E(k,tau)} with respect to the time lag $\tau$ yields the wavenumber--frequency spectrum of passive scalars, 
\begin{equation}
\begin{aligned}
\Phi(k, \omega)
&= \frac{1}{2\pi} \int \mathrm{d}\tau  E(\bb{k}, \tau) e^{i\omega\tau} \\
&= \frac{E(\bb{k})}{\sqrt{2\pi k^2 V^2}}
\exp\left[-\frac{(\omega - \mathbf{k} \cdot \mathbf{U})^2}{2k^2 V^2}\right],
\end{aligned}
\end{equation}
which exhibits a Doppler-like frequency shift $\omega_0=\mathbf{k}\cdot\mathbf{U}$ and Gaussian frequency broadening associated with random sweeping by turbulent fluctuations.

The present model generalizes the Kraichnan model by replacing white-noise advection with colored-noise advection. In the white-noise model, large eddies possess vanishing correlation time, producing purely exponential decay of the temporal correlations of scalar modes. By contrast, the generalized model ensures that the temporal correlations are dominated by the slow dynamics of mean flow and energy-containing large-scale eddies. This faithfully captures the sweeping mechanism of small-scale scalar structures by these large eddies~\cite{kraichnan1964kolmogorov,tennekes1975eulerian}, thereby reproducing the Gaussian temporal decorrelation observed in realistic scalar turbulence~\cite{yeung2002random,Gorbunova_2021,canet2022functional}.

\subsection{Space correlation}
The closed-form equation for the equal-time spatial correlation is
derived from the advection--diffusion equation. For Gaussian velocity statistics, the turbulent transport flux can be expressed by the FND theorem as a memory integral. The velocity-increment covariance confines the relevant memory delays to
$\sigma\lesssim\tau_c(r)$. Therefore, when $\tau_c(r)\ll t_r$, the
response function admits a short-memory expansion analogous
to that introduced by Chertkov \textit{et al.}~\cite{Chertkov1996}. Retaining the zeroth-order response yields the following approximate closed equal-time correlation equation
(see Appendix~\ref{app:SC} for details):

\begin{equation}\label{eq:SC}
\frac{\partial}{\partial r_i} \left[ S_{ij}\, \frac{\partial C({\bf r},0)}{\partial r_j} \right]
+ \chi\left(\frac{r}{L}\right)=0 ,
\end{equation}
where the effective two-particle diffusivity is
\begin{equation}
S_{ij}(\mathbf r)\simeq 2\kappa\delta_{ij}+S_{ij}^{(0)}(\mathbf r),
\end{equation}
and the zeroth-order turbulent relative-diffusion tensor is
\begin{equation}
S_{ij}^{(0)}(\mathbf r)
=\int_0^\infty \mathrm{d}\sigma\,
\left\langle\delta u_i(\mathbf r,t)\delta u_j(\mathbf r,t-\sigma)\right\rangle.
\end{equation}
Here $\delta u_i(\mathbf r,t)=u_i(\mathbf x+\mathbf r,t)-u_i(\mathbf x,t)$.
Equation~\eqref{eq:SC} expresses the balance between turbulent diffusion and large-scale scalar-variance injection at statistical steady state. For the covariance in Eq.~\eqref{eq:def-model}, the effective diffusivity is
\begin{equation}\label{eq:S}
\begin{aligned}
S_{ij}(\mathbf r)
&=2D_3^{-1}\,d_{ij}^{(\zeta)}(\mathbf r)+2\kappa\delta_{ij}\\
&=2D_3^{-1}D_1^{(\zeta)}
\left[(2+\zeta)\delta_{ij}-\zeta\frac{r_i r_j}{r^2}\right]r^\zeta
+2\kappa\delta_{ij},
\end{aligned}
\end{equation}
where $\zeta=\alpha+\beta$. Thus the result inherits both the spatial
and temporal scaling exponents of the velocity covariance.

For isotropic velocity statistics, the equation for space correlation~\eqref{eq:SC} reduces to the radial form
\begin{equation}\label{eq:SC-radial}
\frac{1}{r^2} \frac{\partial}{\partial r} \left( r^2 S_{||}(r) \frac{\partial C(r,0)}{\partial r} \right) + \chi(0) = 0,
\end{equation}
with the longitudinal two-particle diffusivity given by
\begin{equation}\label{eq:S||}
S_{||}(r) = \frac{r_i r_j S_{ij}({\bf r})}{r^2} = 4D_3^{-1}D_1^{(\zeta)}r^{\zeta}+2\kappa.
\end{equation}
Solving Eqs.~\eqref{eq:SC-radial} and \eqref{eq:S||} yields a self-similar spatial correlation of the scalar field in the inertial-convective subrange $r_d \equiv [\kappa D_3/(D_1^{(\zeta)})]^{1/\zeta} \ll r \ll L$,
\begin{equation}\label{eq:space-corraltion-sol}
C({\bf r},0) = \langle \theta^2 \rangle
- \frac{D_3 \chi(0)}{12 D_1^{(\zeta)} (2-\zeta)}\, r^{2-\zeta}, 
\qquad \zeta = \alpha + \beta,
\end{equation}
where $\langle \theta^2 \rangle$ is the mean-squared scalar fluctuation. 

As a result, the scalar structure function $
d_{\theta}({\bf r})
\equiv
\big\langle
[\theta({\bf x}+{\bf r}) - \theta({\bf x})]^2
\big\rangle$ scales as $  d_{\theta}(\mathbf{r})\propto r^{2-\zeta}$. For Kolmogorov velocity scaling exponents ($\alpha = \beta = 2/3$), $\zeta = 4/3$, recovering the classical 
Obukhov--Corrsin $2/3$ law. This indicates that the spatial decorrelation of the scalar field is dominated by distortion from eddies at the corresponding scale in the inertial subrange. In contrast, in the Kraichnan white-noise model, the velocity field is $\delta$-correlated in time, the diffusivity tensor scales as
\(
S_{ij}({\bf r}) \sim r^{\alpha},
\)
leading to
\(
d_{\theta}({\bf r}) \sim r^{2-\alpha}.
\) Although this idealized model serves as a powerful theoretical tool for understanding anomalous scaling, its white-noise approximation does not account for the scale-dependent correlation times of realistic turbulence. Consequently, this leads to a departure from the matching spatial scaling exponents of the velocity and scalar fields (both $2/3$) expected in the inertial-convective subrange.

A dimensional argument for the diffusivity scaling in Eqs.~\eqref{eq:S} and~\eqref{eq:S||} is provided by a scaling-consistency analysis~\cite{chaves2003lagrangian}. The colored-noise model is formally scale invariant under $r\to\lambda r$ and $t\to\lambda^z t$ with $z>0$, so a single characteristic scale $r$ governs the inertial-convective subrange, allowing a simple dimensional argument. With $\delta u(r)\sim r^{\alpha/2}$ and $\tau_c(r)\sim r^\beta$, the eddy turnover time scales as $t_r\sim r/\delta u(r)\sim r^{1-\alpha/2}$. Requiring the two characteristic times to have the same scale dependence gives $\alpha+2\beta=2$, under which
\begin{equation}
    S_{\parallel}(r) \sim \langle\delta u^2(r)\rangle \tau_c(r) \sim r^{\alpha+\beta} \equiv r^\zeta.
\end{equation}
For Kolmogorov scaling ($\alpha=\beta=2/3$), this condition is satisfied, and thus the diffusivity scalings in Eqs.~\eqref{eq:S} and~\eqref{eq:S||} are easily obtained. 

\add{It should be noted, however, that this scaling relation by itself does not establish a parametrically controlled short-memory expansion. Since
$$
\delta u_r\sim \sqrt{D_1^{(\alpha)}}\,r^{\alpha/2}~\sim  \sqrt{D_2}\,r^{\alpha/2},
$$
the corresponding Kubo number is
$$
\mathrm{Ku}(r)\equiv \frac{\tau_c(r)}{t_r}
\equiv \frac{\tau_c(r)}{r/\delta u_r}
\sim
\frac{\sqrt{D_2}}{D_3}\,
r^{\alpha/2+\beta-1}.
$$
Thus, in the Kolmogorov case, \(\alpha+2\beta=2\), the Kubo number is scale independent, and the zeroth-order short-memory closure is controlled only when
$$
\mathrm{Ku}\sim \frac{\sqrt{D_2}}{D_3}\ll 1.
$$
This condition is satisfied by taking \(D_3\) sufficiently large while holding \(D_2/D_3\) fixed. This corresponds to sufficiently rapid temporal decorrelation and hence to a small Kubo number. The resulting diffusivity scalings should therefore be regarded as leading-order closure predictions.}

For the infrared-regularized colored-noise model considered here, the 
finite-correlation-time expansion of the multipoint scalar operator governing 
space correlations introduces a logarithmic scale dependence correction to the zeroth-order expansion. On the 
marginal line \(\alpha+2\beta=2\), the first temporal moment entering this 
correction contains a contribution proportional to \(r^2\ln(L/r)\). As a 
result, the expansion parameter becomes scale dependent, preventing the 
determination of a single scale-independent correction to the zero-order anomalous exponent for higher-order 
scalar structure functions within the present framework. For exactly self-similar colored-noise ensembles, the 
corresponding expansion parameter remains scale independent and anomalous 
exponents can be obtained perturbatively~\cite{Chertkov1996}.

\section{Elliptic approximation to space--time correlations}\label{sec:EA}

Substituting the space correlation~\eqref{eq:space-corraltion-sol} into the Gaussian convolution~\eqref{eq:R=R_sP}, yields the space--time correlation in the inertial-convective subrange,
\begin{equation}\label{eq:exact-sol}
\begin{aligned}
C(\mathbf{r}, \tau) 
= \langle \theta^2 \rangle
 - & \frac{D_3 \, \chi(0)}{12 \, D_1^{(\zeta)} (2-\zeta)} \, (C^{(\zeta)} V \tau)^{2-\zeta} \\
& \times
{}_1F_1\Bigg( \frac{\zeta}{2} - 1, \frac{3}{2}, - \frac{(r-U\tau)^2}{2 V^2 \tau^2} \Bigg),
\end{aligned}
\end{equation}
where ${}_1F_1(a;b;x)$ is the Kummer confluent hypergeometric function
\[{}_1F_1(a;b;x) = \frac{\Gamma(b)}{\Gamma(b-a)\Gamma(a)}
\int_0^1 dt~ e^{x t} t^{a-1} (1-t)^{b-a-1},\] and $ 
\Gamma(x) = \int_0^\infty dt~ t^{x-1} e^{-t}$ is the Gamma function. The dimensionless coefficient $C^{(\zeta)}$ depends solely on the velocity scaling exponent $\zeta = \alpha + \beta$,
\begin{equation}
{C^{(\zeta)}} = \sqrt 2 {\left( {\frac{2}{{\sqrt \pi  }}\Gamma \left( {\frac{{5 - \zeta }}{2}} \right)} \right)^{\frac{1}{{2 - \zeta }}}}.
\end{equation}
For Obukhov--Corrsin scaling of scalar fluctuations ($\zeta=4/3$), $C^{(\zeta)}=1.55$. This value is a prediction of the present closure model. It is determined by the effective exponent \(\zeta\) governing the scalar spatial correlation and by the Gaussian form of scalar temporal decorrelation, rather than by the detailed shape of the colored-noise
kernel. The exponential temporal kernel can be replaced by any normalized integrable correlation kernel \(\rho(s)\) that decays to zero as \(s\rightarrow\infty\), with the same correlation-time scaling \(\tau_c(r)\sim r^\beta\). It follows directly from Eq.~\eqref{eq:exact-sol} that the spatio-temporal structure function $d_{\theta}(r,\tau)=\langle (\theta(\bb{x}+r\bb{e}_x,t+\tau)-\theta(\bb{x},t))^2 \rangle$ is scale-invariant,
\begin{equation}
d_{\theta}(\lambda r,\lambda \tau)=\lambda^{2-\zeta}d_{\theta}(r,\tau),
\end{equation} 
which implies the self-similar isocorrelation contours in the $(r,\tau)$ plane.

\begin{figure}[htbp]
	\centering
	\includegraphics[width=\linewidth]{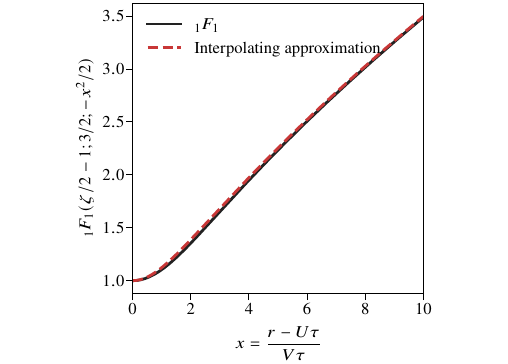}
 \caption{Comparison between the exact Kummer confluent hypergeometric function 
${}_1F_1\!\left(\dfrac{\zeta}{2}-1;\dfrac{3}{2};-\dfrac{(r-U\tau)^2}{2V^2\tau^2}\right)$
(solid line) and the interpolating approximation 
$\left(1+\left(\dfrac{r-U\tau}{C^{(\zeta)}V\tau}\right)^2\right)^{(2-\zeta)/2}$ (dashed line) 
for $\zeta=4/3$. 
The approximation reproduces the correct asymptotic behavior in the limits of 
$\frac{r-U\tau}{V\tau}\to 0$ and $\frac{r-U\tau}{V\tau}\to\infty$.
}
\label{fig:hypergeom_interp}
\end{figure}

The space--time correlation can be uniformly approximated by the following expression that interpolates between the small and large limits of the spatial-to-temporal separation ratio $\frac{r-U\tau}{V\tau}$ in the co-moving frame,
\begin{equation}\label{eq:EA-formula}
C(\mathbf{r},\tau)=\langle\theta^2\rangle-\frac{D_3\chi(0)}{12D_1^{(\zeta)}(2-\zeta)}\big[(r-U\tau)^2+(C^{(\zeta)}V\tau)^2\big]^{\frac{2-\zeta}{2}}.
\end{equation}
This approximation is obtained by replacing the Kummer confluent hypergeometric function with the interpolating form
\begin{align}\label{eq:asym-approx}
    {}_1F_1\Big(\frac{\zeta}{2}-1;\frac{3}{2};-\frac{(r-U\tau)^2}{2 V^2 \tau^2}\Big) \simeq \left(1 + \left
(\frac{r-U\tau}{C^{(\zeta)} V \tau}\right)^2\right)^{(2-\zeta)/2},
\end{align}
which is constructed to reproduce the asymptotic behavior of the function in the limits of
 $\frac{r-U\tau}{V\tau} \to 0$ and $\frac{r-U\tau}{V\tau} \to \infty$~\cite{olver1997asymptotics}. Figure~\ref{fig:hypergeom_interp} compares the exact hypergeometric function with the interpolating approximation, demonstrating excellent agreement across the entire range of $(r-U\tau)/(V\tau)$.

\begin{figure}[!htbp]
	\centering
	\includegraphics[width=\linewidth]{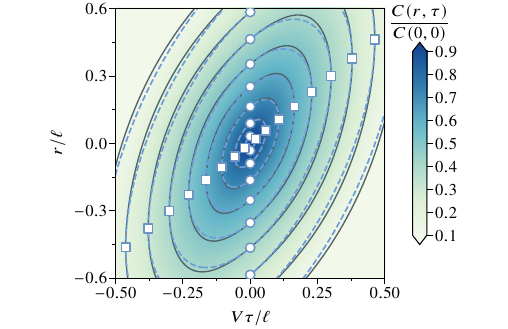}
	\caption{Space--time correlations of passive scalars in the inertial-convective subrange for \(U/V=1\) and \(\zeta=4/3\). The characteristic length is defined as
\(
\ell\equiv
\left[
\cfrac{12D_1^{(\zeta)}(2-\zeta)\langle\theta^2\rangle}
{D_3\chi(0)}
\right]^{1/(2-\zeta)}.
\)
The color map and solid contour lines show the exact solution, Eq.~\eqref{eq:exact-sol}, while the dashed lines denote the iso-correlation contours predicted by the EA model, Eq.~\eqref{eq:EA-model}. Both are obtained for the scalar dynamics governed by Eq.~\eqref{eq:dyn} with the same colored-noise velocity field defined in Eq.~\eqref{eq:def-model}. Open circles and squares indicate the points at which the exact and EA contours coincide along \(\tau=0\) and \(r=U\tau\), respectively.
}
    \label{fig:space--time-corr}
\end{figure}

The elliptic approximation (EA) model follows directly from Eq.~\eqref{eq:EA-formula}:
\begin{equation}\label{eq:EA-model}
C(\mathbf{r},\tau)=C\left(\sqrt{(r-U\tau)^2+(V_{\rm EA}\tau)^2},0\right),\quad V_{\rm EA}=C^{(\zeta)}V,
\end{equation}
where \(V_{\rm EA}\) accounts for the combined effects of large-scale sweeping and small-scale distortion. Both the approximate expression~\eqref{eq:EA-formula} and the EA model are scale-invariant, yielding self-similar space--time correlation contours. For scalar fluctuations obeying Obukhov--Corrsin scaling, \(C^{(\zeta)}=1.55\). Thus, in the co-moving coordinates \((r-U\tau,V\tau)\), the elliptical contours have a constant spatial-to-temporal intercept ratio of \(1.55\). Figure~\ref{fig:space--time-corr} shows the space–time correlation in the $(r,\tau)$ plane. Colored contours with black lines correspond to the exact solution, while the red dashed curves represent the iso-correlation lines predicted by the elliptic approximation. The close agreement between them confirms that the EA model accurately captures the scale-invariance of the space--time correlation.

Although the EA model for passive scalars is derived in this work for a colored-noise velocity field under several approximations, its elliptic self-similar form has been confirmed experimentally in turbulent Rayleigh–B\'enard convection~\cite{he2010small} and in turbulent pipe flow~\cite{Huixin_2025,Li_2026}. However, the universality of the predicted constant $C^{(\zeta)} = 1.55$ remains to be validated by independent direct numerical simulations (DNS) or experimental passive-scalar data. Within the present framework, this value is expected to apply in the inertial-convective range for $Sc = 1$ at sufficiently high Reynolds numbers ($Re$), and may extend to $Sc \neq 1$ when the space--time correlation remain dominated by the inertial-convective range. For $Sc\gg1$ or $Sc\ll1$, additional scalar regimes (such as the viscous-convective or inertial-diffusive ranges) emerge, which may alter the scaling behavior of the correlation.

The space--time correlation of a passive scalar advected by a colored-noise velocity field provides a unified physical picture of scalar decorrelation. The evolution of a scalar element can be decomposed into three concurrent mechanisms: (i) advection by the mean flow, (ii) sweeping by large-scale eddies, and (iii) distortion by small-scale straining. Temporal decorrelation is dominated by the first two processes (mean advection and random sweeping), whereas spatial decorrelation is controlled by the third one (small-scale distortion). Together, these mechanisms furnish the physical foundation of the elliptic approximation~\eqref{eq:EA-model}, consistent with the picture proposed in the TK model~\cite{TK_wang2025}. 

The TK model already derives the elliptic scalar space--time correlation analytically and attributes temporal decorrelation to mean advection and random sweeping, thereby confirming the EA model. The present model instead represents the fluctuating velocity as a single Gaussian colored-noise field with $\tau_c(k)\sim k^{-\beta}$, avoiding a prescribed decomposition into frozen-sweeping and white-noise components. It further predicts that equal-time scalar spatial scaling is governed by $\zeta=\alpha+\beta$, recovering the Obukhov--Corrsin scaling for $\alpha=\beta=2/3$, while generating Gaussian temporal decorrelation dynamically.
 
\section{Conclusions}\label{sec:conclusion}

In this work, we investigate the space--time correlations of passive scalars advected by a colored-noise velocity field with a prescribed power-law spatial spectrum and wavenumber-dependent finite correlation time. We derive analytical expressions for the space--time correlation within the inertial-convective subrange. Our analytical results validate the EA model: in the co-moving space--time frame $(r-U\tau, V\tau)$, the iso-correlation contours are self-similar, and the ratio of spatial to temporal intercepts yields a constant of 1.55. \add{These results are obtained within the leading-order short-memory closure and are quantitatively controlled when the Kubo number is small.}

The colored-noise model accounts for the influence of the velocity correlation timescale on scalar correlations in the inertial-convective subrange. For spatial correlations, it successfully recovers the Obukhov--Corrsin scaling given the advecting velocity with Kolmogorov scaling. In contrast, the white-noise model cannot satisfy both scalings simultaneously. For time correlations, the colored-noise model faithfully reproduces the random-sweeping decorrelation mechanism, yielding a Gaussian decay for the time correlations of scalar modes. This behavior is consistent with observations of scalars in realistic turbulence, whereas the white-noise model predicts an exponential time decorrelation.

The colored-noise model provides a theoretical foundation for understanding the coupling between spatial and temporal decorrelations in scalar turbulence. It is used to clarify the underlying mechanism of the scalar decorrelation process: mean-flow advection and large-eddy sweeping dominate temporal correlation, and small-eddy distortion determines spatial correlation, which has been proposed by the TK model. This fundamental understanding offers crucial insights for developing advanced models of scalar mixing and turbulent transport, such as time-accurate large-eddy simulations~\cite{he2017space} and resolvent analysis~\cite{Wu_He_2023,Wu_Zhang_He_2025}.
\begin{acknowledgments}
This work was supported by the National Natural Science Foundation of China (NSFC) through the Excellence Research Group Program for Multiscale Problems in Nonlinear Mechanics (Grant No.~12588201). 
\end{acknowledgments}

%%%%%%%%%%%%%%%%%%%%|Appendix Begin|%%%%%%%%%%%%%%%%%%%%
\appendix
\section{Derivation of the Space--Time Correlation Equation}
\label{app:STC}

We derive the approximate closed equation for
\begin{equation}
C(\mathbf r,\tau)
=
\left\langle
\theta(\mathbf x,t)
\theta(\mathbf x+\mathbf r,t+\tau)
\right\rangle ,
\qquad \tau>0 .
\end{equation}
For compactness, we introduce
\(
\theta^+=\theta(\mathbf x+\mathbf r,t+\tau)
\),
\(u_i^+=u_i(\mathbf x+\mathbf r,t+\tau)\),
and
\(
\theta^0=\theta(\mathbf x,t)
\).
Differentiation with respect to $\tau$ gives
\begin{align}
\partial_\tau C
=
-U_i\partial_{r_i}C
+\kappa\nabla_{\mathbf r}^{2}C
-\partial_{r_i}{\cal F}_i ,
\label{app-a:STCE}
\end{align}
where the forcing contribution vanishes for \(\tau>0\) because the
forcing is assumed to be temporally white and statistically independent
of the velocity field, and
\begin{equation}
{\cal F}_i(\mathbf r,\tau)
=
\left\langle
u_i^+\theta^+\theta^0
\right\rangle .
\end{equation}
For Gaussian velocity statistics, the FND theorem
gives
\begin{align}
{\cal F}_i
=&
\int_{-\infty}^{t+\tau}\!ds
\int d\mathbf z\,
D_{ij}(\mathbf x+\mathbf r-\mathbf z,t+\tau-s)
\left\langle
\frac{\delta(\theta^+\theta^0)}
{\delta u_j(\mathbf z,s)}
\right\rangle .
\label{app-a:scalar-flux}
\end{align}
We split this exact response integral into two temporal regions:
the recent contribution \({\cal F}_i^{(A)}\), corresponding to
\(t<s<t+\tau\), and the pre-\(t\) memory contribution
\({\cal F}_i^{(B)}\), corresponding to \(s<t\):
\begin{equation}
{\cal F}_i
=
{\cal F}_i^{(A)}
+
{\cal F}_i^{(B)} .
\end{equation}
The closure approximation is introduced below when evaluating the
response functions in the two temporal intervals, under the short-memory
condition
\(
\tau_c(r)\ll t_r
\),
where \(\tau_c(r)\) denotes the velocity correlation time and \(t_r\)
the characteristic turnover time at scale \(r\).

For the time interval $t<s<t+\tau$, the velocity perturbation
affects only the scalar field at the advanced time $t+\tau$.
The corresponding response function is
\begin{equation}
\frac{\delta\theta^+}{\delta u_j(\mathbf z,s)}
=
-G_{\kappa}
(\mathbf x+\mathbf r,t+\tau;\mathbf z,s)
\partial_{z_j}\theta(\mathbf z,s).
\end{equation}
Introducing the backward time delay
\(
\sigma=t+\tau-s
\),
we perform a short-time expansion of the response propagator. The
expansion is justified when the time lag $\sigma<\tau$ is much shorter than the turnover time $t_r$. Under this condition, the
propagator can be written as
\begin{align}
G_{\kappa}
(\mathbf y^+,t^+;\mathbf z,t^+-\sigma)
={}&
\delta^{(3)}(\mathbf y^+-\mathbf z)
\nonumber\\
&+\sigma{\cal L}_{\mathbf z}(t^+)
\delta^{(3)}(\mathbf y^+-\mathbf z)
+\mathcal O\left(\sigma^2\right),
\end{align}
where
$
{\cal L}_{\mathbf z}(t)
=
\kappa\nabla_z^2
-\bigl(U_k+u_k(\mathbf z,t)\bigr)\partial_{z_k}
$
denotes the instantaneous advection--diffusion operator. The scalar
gradient appearing in the response function has the corresponding
short-time expansion
\begin{align}
\partial_{z_j}\theta(\mathbf z,t^+-\sigma)
=
\partial_{z_j}\theta(\mathbf z,t^+)
-\sigma\partial_t\partial_{z_j}\theta(\mathbf z,t^+)
+\mathcal O\left(\sigma^2\right).
\end{align}
Keeping only the leading-order contribution in this expansion gives
\begin{equation}
{\cal F}_i^{(A)}
\simeq
-K_{ij}^{(0)}(\tau)\partial_{r_j}C,
\end{equation}
with
\begin{equation}
K_{ij}^{(0)}(\tau)
=
\int_0^\tau
D_{ij}(\mathbf0,\sigma)d\sigma.
\label{app-a:Kij-SCTE}
\end{equation}

The contribution from the velocity history prior to time \(t\) is given by
\begin{equation}
\mathcal{F}_i^{(B)}
=
\int_{-\infty}^{t} \mathrm{d}s
\int \mathrm{d}\mathbf{z}\,
D_{ij}(\mathbf{x}+\mathbf{r}-\mathbf{z},t+\tau-s)
\left\langle
\frac{\delta(\theta^+\theta^0)}
{\delta u_j(\mathbf{z},s)}
\right\rangle.
\end{equation}
By introducing the variables \(q = t - s \geq 0\) and \(\boldsymbol{\rho} = \mathbf{z} - \mathbf{x}\), this contribution reduces to
\begin{equation}
\mathcal{F}_i^{(B)}
=
\int_0^\infty \mathrm{d}q
\int \mathrm{d}\boldsymbol{\rho}\,
D_{ij}(\mathbf{r}-\boldsymbol{\rho},\tau+q)
\mathcal{R}_j(\boldsymbol{\rho},q),
\end{equation}
where \(\mathcal{R}_j\) denotes the corresponding averaged response function
\[
\mathcal{R}_j
=
\left\langle
\theta^0
\frac{\delta\theta^+}
{\delta u_j(\mathbf{x}+\boldsymbol{\rho},t-q)}
\right\rangle
+
\left\langle
\theta^+
\frac{\delta\theta^0}
{\delta u_j(\mathbf{x}+\boldsymbol{\rho},t-q)}
\right\rangle .
\]
The convolution can be expressed using the Fourier form
\begin{equation}
\mathcal{F}_i^{(B)}
=
\int_0^\infty \mathrm{d}q
\int \frac{\mathrm{d}\mathbf{k}}{(2\pi)^3}\,
e^{i\mathbf{k}\cdot\mathbf{r}}
\widehat{D}_{ij}(\mathbf{k},\tau+q)
\widehat{\mathcal{R}}_j(\mathbf{k},q).
\end{equation}
For the colored-noise velocity model,
\begin{equation}
\widehat{D}_{ij}(\mathbf{k},\tau+q)
\propto
\exp\!\left[-D_3 k_L^\beta (\tau+q)\right].
\end{equation}
Hence, each Fourier mode carries the memory factor
\begin{equation}
e^{-\tau/\tau_k},
\qquad
\tau_k = (D_3 k_L^\beta)^{-1}.
\end{equation}
Therefore, the pre-\(t\) contribution is exponentially suppressed
for \(\tau\gg\tau_c(r)\sim D_3^{-1}r^{\beta}\). In the limit of white-noise velocity \(\tau_c(r) \to 0\), this prior-history contribution vanishes identically. Therefore, within the short-memory approximation adopted here, 
\(\mathcal F_i^{(B)}\) is neglected as an additional short-memory closure approximation.

Combining the two response intervals, the turbulent scalar flux is
approximated as
\begin{equation}
{\cal F}_i
\simeq
-K_{ij}^{(0)}(\tau)\partial_{r_j}C .
\end{equation}
Substituting this closure into Eq.~\eqref{app-a:STCE} gives the
space--time correlation equation
\begin{equation}
\frac{\partial C({\bf r},\tau)}{\partial \tau}
=
-U_i\frac{\partial C({\bf r},\tau)}{\partial r_i}
+\frac{\partial}{\partial r_i}
\left[
K_{ij}(\tau)
\frac{\partial C({\bf r},\tau)}{\partial r_j}
\right],
\end{equation}
where
$
K_{ij}(\tau)
\simeq
\kappa\delta_{ij}+K_{ij}^{(0)}(\tau).
$

\section{Derivation of the Space Correlation Equation}
\label{app:SC}
We next derive the approximate closed equation for the equal-time correlation
\begin{equation}
C(\mathbf r,0)
=
\left\langle
\theta(\mathbf x+\mathbf r,t)
\theta(\mathbf x,t)
\right\rangle .
\end{equation}
For convenience, we introduce the notation
\(
\theta^r=\theta(\mathbf x+\mathbf r,t)
\)
and
\(
\theta^0=\theta(\mathbf x,t)
\).
The velocity increment across the separation \(\mathbf r\) is defined as
\begin{equation}
\delta u_i(\mathbf r,t)
=
u_i(\mathbf x+\mathbf r,t)-u_i(\mathbf x,t).
\end{equation}

Using incompressibility, statistical homogeneity, and stationarity of
the scalar field, the two-point balance equation can be written as
\begin{equation}
2\kappa\nabla_{\mathbf r}^{2}C
-\partial_{r_i}{\cal F}_i
+\chi(r/L)=0 ,
\label{app-b:SCE}
\end{equation}
where the scalar flux associated with the velocity increment is defined
as
\begin{equation}
{\cal F}_i(\mathbf r)
=
\left\langle
\delta u_i(\mathbf r,t)
\theta^r\theta^0
\right\rangle .
\end{equation}

For a Gaussian velocity field, the FND theorem allows the scalar flux to be expressed
in terms of the velocity covariance and the functional response of the
scalar field:
\begin{align}
{\cal F}_i
=&
\int_{-\infty}^{t}\!ds
\int d\mathbf z\,
\left\langle
\delta u_i(\mathbf r,t)
u_j(\mathbf z,s)
\right\rangle
\left\langle
\frac{\delta(\theta^r\theta^0)}
{\delta u_j(\mathbf z,s)}
\right\rangle .
\label{app-b-flux}
\end{align}
Introducing the backward time delay
\(
\sigma=t-s
\),
the exact response of the scalar pair to a velocity perturbation at the
earlier time \(t-\sigma\) is
\begin{align}
\frac{\delta(\theta^r\theta^0)}
{\delta u_j(\mathbf z,t-\sigma)}
=&
-G_{\kappa}
(\mathbf x+\mathbf r,t;\mathbf z,t-\sigma)
\partial_{z_j}\theta(\mathbf z,t-\sigma)\theta^0
\nonumber\\
&-
G_{\kappa}
(\mathbf x,t;\mathbf z,t-\sigma)
\theta^r
\partial_{z_j}\theta(\mathbf z,t-\sigma).
\label{app-b:response-exact}
\end{align}

We apply the short-time expansion to the response function in
Eq.~\eqref{app-b:response-exact}. For the colored-noise velocity
field considered here, the contribution from large time delays is
suppressed by the decay of the velocity correlation, and the dominant
response originates from the interval
\(
\sigma\lesssim\tau_c(r)
\).
When the velocity correlation time at scale \(r\) is much shorter than the characteristic turnover time at the same scale,
\(
\tau_c(r)\ll t_r
\),
where \(t_r\) denotes the turnover time at scale \(r\), the response propagator can be expanded as
\begin{align}
G_{\kappa}
(\mathbf y,t;\mathbf z,t-\sigma)
=&
\delta^{(3)}(\mathbf y-\mathbf z)
\nonumber\\
&+
\sigma{\cal L}_{\mathbf z}(t)
\delta^{(3)}(\mathbf y-\mathbf z)
+\mathcal O\left(\sigma^2\right).
\end{align}
The scalar
gradient appearing in the response function has the corresponding
short-time expansion
\begin{align}
\partial_{z_j}\theta(\mathbf z,t-\sigma)
=
\partial_{z_j}\theta(\mathbf z,t)
-
\sigma
\partial_t\partial_{z_j}\theta(\mathbf z,t)
+\mathcal O\left(\sigma^2\right).
\end{align}
Retaining only the leading-order term in the short-memory expansion yields the zeroth-order closure for the scalar flux:
\begin{equation}
{\cal F}_i
\simeq
-
S_{ij}^{(0)}(\mathbf r)
\partial_{r_j}C .
\label{eq:SC-flux-final}
\end{equation}
where the integrated velocity-increment covariance tensor is defined as
\begin{equation}
\begin{aligned}
S_{ij}^{(0)}(\mathbf r)
&\equiv
\int_0^\infty d\sigma\,
\left\langle
\delta u_i(\mathbf r,t)
\delta u_j(\mathbf r,t-\sigma)
\right\rangle
\\&
=
2\int_0^\infty d\sigma
\left[
D_{ij}(\mathbf0,\sigma)
-
D_{ij}(\mathbf r,\sigma)
\right].
\end{aligned}
\label{eq:SC-S0}
\end{equation}

Finally, substituting the flux closure from
Eq.~\eqref{eq:SC-flux-final} into the equal-time balance equation in
Eq.~\eqref{app-b:SCE} yields the closed equation for the equal-time
scalar correlation:
\begin{equation}
\frac{\partial}{\partial r_i} \left[ S_{ij}\, \frac{\partial C({\bf r},0)}{\partial r_j} \right]
+ \chi\left(\frac{r}{L}\right)=0 ,
\end{equation}
where $
S_{ij}(\mathbf r)\simeq 2\kappa\delta_{ij}+S_{ij}^{(0)}(\mathbf r)$.

%%%%%%%%%%%%%%%%|Appendix End|%%%%%%%%%%%%%%%%%%%%
%\end{CJK}
% \bibliographystyle{apsrev4-2}
\bibliography{reference}

\end{document}